\documentclass[twocolumn,aps,superscriptaddress,prl]{revtex4-1}
\usepackage{graphicx}
\usepackage{amssymb}
\usepackage{amsmath}
\usepackage{bm}
\usepackage{color}

\usepackage{braket}
\usepackage{siunitx}
\usepackage[top=30truemm,bottom=30truemm,left=25truemm,right=25truemm]{geometry}

\def\Vec{\mathbf}

\def\lsim{\, \lower -0.3ex \hbox{$<$} \kern -0.75em \lower 0.7ex \hbox{$\sim$} \,}
\def\gsim{\, \lower -0.3ex \hbox{$>$} \kern -0.75em \lower 0.7ex \hbox{$\sim$} \,}

\begin{document}

\title{Topological charge pumping by sliding moir\'e pattern}
\author{Manato Fujimoto}
\affiliation{Department of Physics, Osaka University,  Osaka 560-0043, Japan}
\author{Henri Koschke}
\affiliation{Department of Physics, Osaka University,  Osaka 560-0043, Japan}
\affiliation{Department of Physics, University of Cologne, 50923 Cologne, Germany}
\author{Mikito Koshino}
\affiliation{Department of Physics, Osaka University,  Osaka 560-0043, Japan}
\date{\today}

\begin{abstract} 
We study the adiabatic topological charge pumping driven by interlayer sliding in the moir\'{e} superlattices.
We show that, when we slide a single layer of the twisted bilayer system relatively to the other,
a moir\'{e} pattern flow and a quantized transport of electrons occurs.
When the Fermi energy is in a spectral gap,
the number of pumped charges in the interlayer sliding process 
is quantized to a sliding Chern number, which obeys a Diophantine equation
analogous to the quantum Hall effect. We apply the argument to the twisted bilayer graphene,
and find that energy gaps above and below the nearly-flat bands has non-zero sliding Chern numbers.
When the Fermi energy is in either of those gaps, the slide-driven topological pumping occurs perpendicularly to the sliding direction.
\end{abstract}

\maketitle
The spiral pump, also known as the Archimedes's skrew, is a mechanical device to pump up water by a rotating spiral blade. 
The machine converts the cyclic motion of blade to a directed motion of water,
and it works no matter how slowly the screw is rotated.
The quantum version of this phenomena is known as the Thouless pumping \cite{thouless1983quantization}, where
an adiabatic, cyclic movement of a periodic potential pumps particles without a bias voltage.
The amount of the transferred particles is precisely quantized and expressed as a topological invariant.
The experimental realization of quantum pumping requires a precise control of the time-dependent periodic potential.
The electron pumping experiments have been performed in various semiconductor-based nanoscale devices
\cite{switkes1999adiabatic,blumenthal2007gigahertz,kaestner2008single}.
More recently, the topological charge pump was realized in ultracold atoms in optical superlattices 
\cite{nakajima2016topological,lohse2016thouless,lu2016geometrical},
and it was also extensively studied in theory
\cite{chiang1998quantum,qian2011quantum,wang2013topological,matsuda2014topological,mei2014topological,wei2015anomalous,yang2018continuously,matsuda2019two}.

\begin{figure}
  \begin{center}
    \leavevmode\includegraphics[width=1. \hsize]{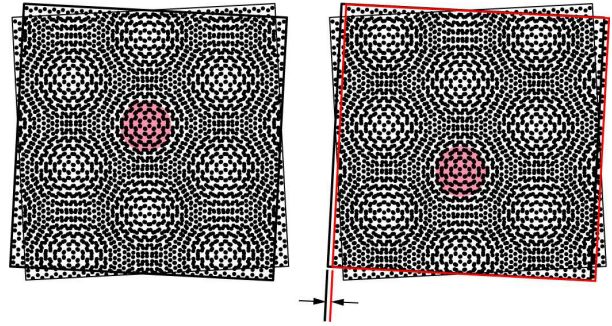}
    \caption{
    Moir\'{e} pumping in twisted bilayer graphene. From left to right, we slide the layer 1 by half of the atomic constant
    and then the moire pattern shifts by the half of the moir\'{e} period as indicated by a pink circle.
    }
    \label{fig_moire_slide}
  \end{center}
  \end{figure}

In this Letter, we propose a simple realization of topological pumping using moir\'{e} superlattice
of the two-dimensional (2D) materials.
When two periodic lattices are overlaid on top of each other with a relative rotation,
the lattice mismatch between the layers gives rise to a moir\'{e} interference pattern.
If we slide a single layer of the twisted bilayer system relatively to the other,
the moir\'{e} pattern flows at a much faster speed 
than the sliding speed as shown in Fig.\ \ref{fig_moire_slide}.
The process is cyclic, as sliding by a single atomic constant shifts the moir\'{e} pattern exactly by a single superlattice period.
We can then ask how the electrons are transported by the movement of the moir\'{e} pattern.

In this paper, we study the topological charge pumping driven by interlayer sliding in the moir\'{e} superlattices.
We first consider a one-dimensional (1D) double-chain model composed of two tight-binding chains with different lattice constants.
We show that the number of pumped charges in the interlayer sliding process
is quantized into a sliding Chern number,
which satisfies a Diophantine equation similar to that for the quantum Hall effect \cite{thouless1982quantized}.
We apply the same argument to the twisted bilayer graphene (TBG)\cite{lopes2007graphene,mele2010commensuration,trambly2010localization,shallcross2010electronic,morell2010flat,bistritzer2011moirepnas,kindermann2011local},
and find that eight Chern numbers are associated with each single gap,
which correspond to different pumping directions under different sliding directions.
In low-angle twisted bilayer graphenes, we show that energy gaps above and below the nearly-flat bands  \cite{bistritzer2011moirepnas,de2012numerical}
has non-zero sliding Chern numbers.
When the Fermi energy in either of those gaps,
the electrons are pumped almost perpendicularly to the sliding direction
following the movement of the moir\'{e} pattern.

\begin{figure}
  \begin{center}
    \leavevmode\includegraphics[width=1. \hsize]{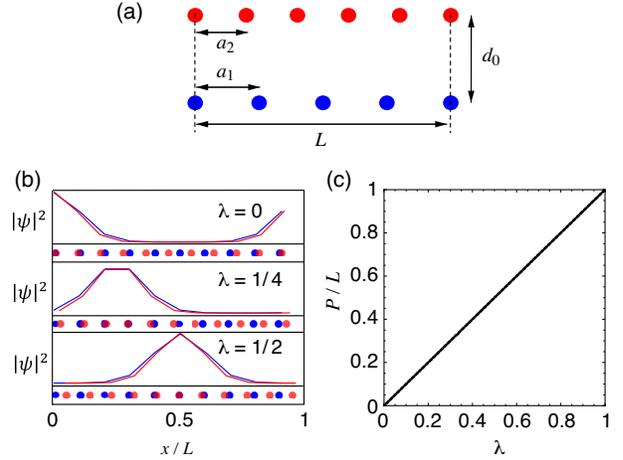}
    \caption{(a) Double chain model with $N_1=4,N_2=5$.
    (b)  Evolution of a wave function of the lowest band in the double chain of $(N_1,N_2)=(10,11)$,
where the chain 2 (red) is shifted by $\lambda a_2$ with the chain 1 (blue) fixed.
    (c) Polarization $P$ [Eq.\ (\ref{eq_P})] in the lowest gap as a function of sliding parameter $\lambda$.
}
    \label{fig_1D}
  \end{center}
  \end{figure}

Let us consider a 1D double chain as illustrated Fig.\ \ref{fig_1D}(a).
The system contains two atomic chains with different lattice constants,
which are arranged parallel to one another in a certain distance $d_0$.
The chains are denoted by $1$ and $2$ and their lattice constants by $a_1$,$a_2$ respectively. 
In the following, we consider a commensurate case $N_1a_1=N_2a_2=L$
with integers $N_1$  and $N_2$,
where an vertical overlap of an atom pair from both chains appears 
in a period $L$. 

 
We calculate the eigenenergies and eigenfunctions
in a double-chain tight-binding model with $s$ atomic orbitals on each single site.
The Hamiltonian is written as  
\begin{equation}
H=-\sum_{\langle i, j\rangle} t\left(\mathbf{R}_{i}-\mathbf{R}_{j}\right)\left|\mathbf{R}_{i}\right\rangle\left\langle\mathbf{R}_{j}\right|+\mathrm{H.c.}
\end{equation}
where $\mathbf{R}_{i}$ and $\left|\mathbf{R}_{i}\right\rangle$ represent the lattice point and the atomic state at site $i$ respectively, and $t\left(\mathbf{R}_{i}-\mathbf{R}_{j}\right)$ is the transfer integral between site $i$ and site $j$. 
For the intra-chain hopping we just take the nearest neighbor hopping 
and it is assumed to be the same for both chains and defined as unit of energy. 
For the hopping between sites on the different chains, we assume
$-t(d)=t_0 e^{-(d-d_0)/\delta_0}$ where $d=|\Vec{R}_{i}-\Vec{R}_{j}|$ and $\delta_0$ is decay length.
In this paper we assume $t_0=4.0,d_0=1.0$ and $\delta_0=0.1$.


We consider an adiabatic charge pumping caused by a relative sliding of chains.
By starting from an initial state in Fig.\ \ref{fig_1D}(a), we horizontally shift either of chain $l=1$ or 2 
by $\lambda a_l  \, (0 \leq \lambda \leq 1)$ with the other chain fixed.
When $\lambda$ is increased from 0 to 1, the Hamiltonian returns to its original state.
We assume that the shift occurs in a sufficiently long time,
so that we can treat the problem as an adiabatic topological pumping \cite{thouless1983quantization}.
The charge transport in such a process is expressed as a change of the polarization.
If the Fermi energy lies inside a certain gap of the spectrum, 
the electric polarization, or the center of mass of the occupied electrons, is given by
\begin{equation}
P(\lambda) =\sum_{n \in \rm occ.}
\frac{L}{2 \pi}  \int_{-\frac{\pi}{L}}^{\frac{\pi}{L}} d k 
\,\,
i\langle u_{nk}(\lambda) | \frac{\partial}{\partial k} | u_{nk}(\lambda)\rangle
\label{eq_P}
\end{equation} 
where $u_{nk}(\lambda)$ is the Bloch eigen state of the $n$th band in the instantaneous Hamiltonian 
at shift $\lambda$,
and occ. represents the occupied bands below the Fermi energy.
The charge transport during the process is then given by $\Delta P= \int_0^1 d\lambda (\partial P / \partial \lambda)$.
This is expressed as $\Delta P=C L$ with the sliding Chern number,
\begin{equation}
C = \sum_{n \in \mathrm{occ.}} 
\frac{iL}{2 \pi} 
\int_{-\frac{\pi}{L}}^{\frac{\pi}{L}} d k \int_{0}^{1} d \lambda
\left[
\Bigl\langle \frac{\partial u}{\partial \lambda} \Bigl| \frac{\partial u }{\partial k} \Bigr\rangle
-
\Bigl\langle \frac{\partial u}{\partial k} \Bigl| \frac{\partial u }{\partial \lambda} \Bigr\rangle
\right],
\label{eq_C}
\end{equation}
where $u=u_{nk}(\lambda)$.
We can define two different Chern numbers  $C_l\, (l=1,2)$ 
for the movement of chain $l$. 

Fig.\ \ref{fig_1D}(b) shows the evolution of a wave function of the lowest band in the double chain of 
$(N_1,N_2)=(10,11)$,
where the chain 2 (red) is shifted by $\lambda a_2$ with the chain 1 (blue) fixed.
The blue and red dots in the bottom represent the horizontal positions of chain 1 and 2, respectively.
We see that the wave center exactly follows the atom overlap region, i.e., the region 
where the chain 1 atoms and chain 2 atoms are overlapping in the horizontal position.
In this particular system $(N_2-N_1=1)$, the center of the overlap region is given by at $x= \lambda L$ 
as a function of sliding paramater $\lambda$,
so it moves exactly by the superlattice period $L$ from $\lambda =0$ to 1.
In Fig.\ \ref{fig_1D}(c), we plot the polarization $P$ [Eq.\ (\ref{eq_P})]
in the lowest gap as a function of shift $\lambda$,
where we actually see that the charge is pumped by $L$ after one cycle, i.e. $C_2=1$.
We can consider a similar process to move chain 1 by fixing chain 2 instead,
and then we have $C_1=-1$, i.e. the change is pumped by a single moir\'{e} period in the negative direction.

\begin{figure}
  \begin{center}
    \leavevmode\includegraphics[width=0.9 \hsize]{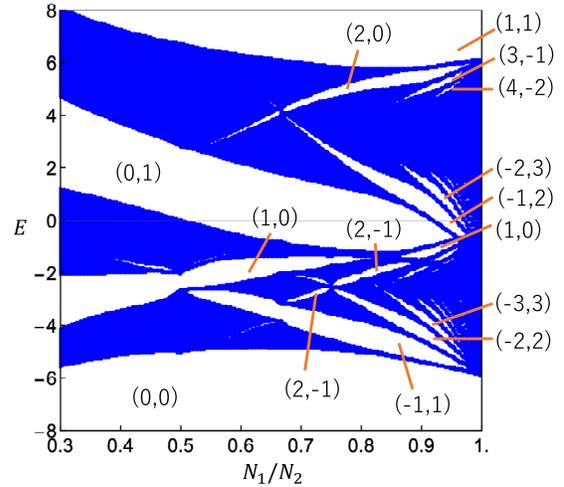}
    \caption{Energy spectrum of the one-dimensional double chain model 
    as a function of $N_1/N_2$. The numbers assigned to gaps indicate the sliding Chern numbers $(C_1,C_2)$. }
    \label{fig_Hof_1D}
  \end{center}
  \end{figure}

We calculate the band structure and the sliding Chern numbers for various configurations of $N_1$ and $N_2$.
In Fig.\ \ref{fig_Hof_1D},  we plot the energy spectrum as a function of $N_1/N_2$,
where the filled area represents the energy region where the eigenstates exist,
and the numbers assigned to the gaps are the sliding Chern numbers $(C_1,C_2)$.
The picture shows some similarities with Hofstadter's butterfly \cite{hofstadter1976energy} 
with the quantized Hall integers
in two-dimensional periodic system under the magnetic field \cite{thouless1982quantized}.
Actually the sliding Chern numbers can be found
by using a Diophantine equation similar to that in the quantum Hall systems \cite{thouless1982quantized},
without integrating the Berry curvature in Eq.\ (\ref{eq_C}), as in the following manner.

Let us consider a double chain specified by $N_1$ and $N_2$,
and assume the Fermi energy lies in a gap with $r$ bands below,
i.e., $r$ bands out of $N_1+N_2$ bands in total are fully occupied.
If we fix chain 2 and shift the chain 1 by $L(=N_1 a_1)$,
the number of pumped electrons is given by $N_1 C_1$
i.e., $N_1 C_1$ electrons passed through any cross section perpendicular to the double chain. 
On the other hand, if we fix chain 1 and shift the chain 2 by $-L(=-N_2 a_2)$,
the number of pumped electrons is given by $-N_2 C_2$,
i.e., $N_2 C_2$ electrons passed in the negative direction.
The former and the latter processes share the same relative motion between the two chains, 
but differ only in the absolute position of the final state by $L$.
If we shift  the whole system (chain 1 and 2 together) by $L$ following the latter process, 
it causes a pump of extra $r$ electrons,
because the number of electrons per a superlattice period $L$ is equal to 
the number of the occupied bands, $r$.
The equality of the two processes leads to $N_1C_1 = - N_2C_2 + r$, or
\begin{equation}
    N_1C_1+N_2C_2=r,
\end{equation}
which is a Diophantine equation for the sliding Chern numbers.
 
If we define the ratio of the double periods as $\alpha = N_1/N_2 = a_2/a_1$
and the electron density as $\rho = r/N_2$, we have $\rho = C_1 \alpha + C_2$.
Now the Chern numbers $C_1$ and $C_2$ can easily be derived from the diagram of Fig.\ \ref{fig_Hof_1D},
by counting the number of states below a particular gap as a function of $\alpha$,
and calculating $\partial \rho / \partial \alpha$. This is an analog to the Str\v{e}da's formula in the integer quantum Hall effect. \cite{streda1982theory}

We can calculate the charge pumping in sliding TBGs in the same manner.
We define the structure of TBG of twist angle $\theta$
by rotating layers 1 and 2 of the AA-stacked bilayer around a center of hexagon by $-\theta/2$ and $+\theta/2$, respectively.
The lattice vectors of layer $l$ are given by $\mathbf{a}_{i}^{(l)}=R(\mp \theta / 2) \mathbf{a}_{i}$ with $\mp$ for $l=1,2$, respectively, where 
$\mathbf{a}_{1}=a(1,0)$ and $\mathbf{a}_{2}=a(1/2, \sqrt{3}/2)$ are the lattice vectors of the AA-stacked bilayer before the rotation,
$a \approx 0.246$ nm  is the lattice constant, and $R(\theta)$ represents the rotation by $\theta$.
When the rotation angle is small, the mismatch between the lattice vectors 
of the two layers gives rise to a long-range moir\'e pattern 
which is ruled by the primitive lattice vectors,\cite{moon2013opticalabsorption}
\begin{align}
\mathbf{L}_{i}^{M} = [R(\theta/2)-R(-\theta/2)]^{-1}\Vec{a}_i\quad (i=1,2).
\end{align}

The atomic structure of TBG is not exactly periodic in general, because the moir\'{e} period is generally 
incommensurate with the underlying atomic lattice.
A commensurate structure with an exact period takes place when the twist angle $\theta$
coincides with the angle between $\mathbf{v}_1=m \mathbf{a}_1+ n \mathbf{a}_2$ and
 $\mathbf{v}_2=n \mathbf{a}_1+ m \mathbf{a}_2$ with certain integers $m$ and $n$.
Then the lattice points $\mathbf{v}_1$ on layer 1 and $\mathbf{v}_2$ on layer 2 of the non-rotated bilayer graphene merge after 
the rotation and  a rigorously periodic structure is obtained.
A lattice vector of the superlattice unit cell is then given by 
$\mathbf{L} = m \mathbf{a}_{1}^{(1)}+n \mathbf{a}_{2}^{(1)}=n \mathbf{a}_{1}^{(2)}+m \mathbf{a}_{2}^{(2)}$.
The twist angle $\theta$ is equal to the angle between $\mathbf{v}_1$ and $\mathbf{v}_2$,
which is $\cos \theta=(1/2)(m^{2}+n^{2}+4 m n)/(m^{2}+n^{2}+m n).$
The 60$^\circ$ rotation of $\mathbf{L}$ also gives a lattice vector because of $C_6$ rotational symmetry.
We can choose two primitive lattice vectors as $\Vec{L}_1 = R(-2\pi/3)\Vec{L}$ and $\Vec{L}_2 = R(-\pi/3)\Vec{L}$, 
which can be written as
\begin{align}
&\mathbf{L}_{1}= n \mathbf{a}_{1}^{(1)}-(m+n) \mathbf{a}_{2}^{(1)}= m \mathbf{a}_{1}^{(2)}-(m+n) \mathbf{a}_{2}^{(2)}\nonumber\\
&\mathbf{L}_{2}= (m+n) \mathbf{a}_{1}^{(1)}-m \mathbf{a}_{2}^{(1)}= (m+n) \mathbf{a}_{1}^{(2)}-n \mathbf{a}_{2}^{(2)}.
\label{eq_L_rigorous}
\end{align}
In this choice, the rigorous period and the moir\'{e} period are simply related by $\mathbf{L}_{i} = |m-n|\mathbf{L}_{i}^{M}$.\cite{moon2013opticalabsorption}

Now let us consider the charge pumping of a commensurate TBG.
We adiabatically slide the layer $l$($=1$ or 2) 
by its own lattice period $\mathbf{a}_{i}^{(l)}$ ($i=1$ or 2)  while the other layer is fixed.
The variation of polarization in this process is written as
\begin{align}
\Delta \mathbf{P} =& C_{i1}^{(l)}\mathbf{L}_{1}+ C_{i2}^{(l)}\mathbf{L}_{2},
\\
C_{ij}^{(l)}=&
\sum_{n={\rm occ.}}
\frac{i S}{(2\pi)^2}
 \int_{\rm BZ} d^2k \int_{0}^{1} d\lambda_i 
 \nonumber\\
 &\quad
 \left[
\Bigl\langle \frac{\partial u}{\partial \lambda_i} \Bigl| \frac{\partial u }{\partial k_j} \Bigr\rangle
-
\Bigl\langle \frac{\partial u}{\partial k_j} \Bigl| \frac{\partial u }{\partial \lambda_i} \Bigr\rangle
\right],
\end{align}
where $u = u_{n\Vec{k}}(\lambda_1,\lambda_2)$ is the Bloch eigenstate in the instantaneous Hamiltonian 
with the layer $l$ shifted by  $\lambda_1 \mathbf{a}_{1}^{(l)}+ \lambda_2 \mathbf{a}_{2}^{(l)}$,
BZ represents the first superlattice Brillouin zone, $S = |\mathbf{L}_{1}\times\mathbf{L}_{2}|$ is the superlattice unit cell area.
$\partial/\partial k_j= (\Vec{G}_j / |\Vec{G}_j|)\cdot\nabla_\Vec{k}$, 
and $\mathbf{G}_{j}$ is the reciprocal lattice vectors satisfying $\mathbf{G}_{i}\cdot\mathbf{L}_{j} = 2\pi \delta_{ij}$.
The sliding Chern number  $C_{ij}^{(l)}$ represents the number of electrons passed through the unit-cell side perpendicular to $\Vec{G}_{j}$ (i.e., the cross section spanned by $\Vec{L}_2$ for $j=1$, and $\Vec{L}_1$ for $j=2$),
during an adiabatic sliding of the layer $l$ by $\mathbf{a}_{i}^{(l)}$.
It is formally similar to, but different from the Chern number for the quantized Hall conductivity \cite{thouless1982quantized},
as it is related to derivative in mechanical interlayer shift.

To obtain the Diophantine equation for the sliding Chern numbers in the TBG, we follow the same steps as in 1D.
We assume that the Fermi energy is in a gap and $r$ bands are fully occupied.
Considering that a shift of layer $1$ by $\mathbf{L}_{1} (= n \mathbf{a}_{1}^{(1)}-(m+n) \mathbf{a}_{2}^{(1)})$
is equivalent to a shift of
layer 2 by $-\mathbf{L}_{1} (= - m \mathbf{a}_{1}^{(2)}+(m+n) \mathbf{a}_{2}^{(2)})$
followed by a shift of whole system by $\mathbf{L}_{1}$,
we obtain 
\begin{align}
\label{eq_Diophantine_TBG1}
& n C_{11}^{(1)} + m C_{11}^{(2)} - (m+n) (C_{21}^{(1)}+C_{21}^{(2)}) =r, \nonumber\\
& n C_{12}^{(1)} + m C_{12}^{(2)} -(m+n) (C_{22}^{(1)}+C_{22}^{(2)}) =0.
\end{align} 
A similar argument for the shift by $\mathbf{L}_{2}$ gives
\begin{align}
\label{eq_Diophantine_TBG2}
& (m+n) (C_{11}^{(1)}+C_{11}^{(2)}) - m C_{21}^{(1)} - n C_{21}^{(2)}=0, \nonumber\\
& (m+n) (C_{12}^{(1)}+C_{12}^{(2)}) - m C_{22}^{(1)} - n C_{22}^{(2)}=r.
\end{align} 


The exact lattice commensurability is not actually important in low twist angles, 
where the physical property is approximately described by the continuum model, 
which is periodic in the moir\'{e} period $\Vec{L}^M_j$.\cite{lopes2007graphene,bistritzer2011moirepnas,kindermann2011local,PhysRevB.86.155449,moon2013opticalabsorption,koshino2015interlayer,koshino2015electronic,weckbecker2016lowenergy}
Indeed, Eqs.\ (\ref{eq_Diophantine_TBG1}) and (\ref{eq_Diophantine_TBG2}) can also be transformed in a continuous form as follows. 
Since the rigorous period $\Vec{L}_j$ is $|m-n|$ times as large as the moir\'{e} period $\Vec{L}^M_j$,
a single continuum band corresponds to $|m-n|^2$ rigorous bands considering the zone folding,
and therefore the number of occupied continuum bands is given by $\tilde{r} = r/|m-n|^2$.
We can also define $\tilde{C}_{ij}^{(l)} = C_{ij}^{(l)}/|m-n|$ as the number of electrons passed through the cross section 
spanned by $\Vec{L}^M_j$ in the adiabatic sliding of the layer $l$ by $\mathbf{a}_{i}^{(l)}$.
Then Eqs. (\ref{eq_Diophantine_TBG1}) and (\ref{eq_Diophantine_TBG2}) become
\begin{align}
&
\frac{\beta+1}{2} \tilde{C}_{11}^{(1)}
+ \frac{\beta-1}{2} \tilde{C}_{11}^{(2)}
-\beta  (\tilde{C}_{21}^{(1)}+\tilde{C}_{21}^{(2)}) 
=\tilde{r},
\nonumber\\
&
\frac{\beta+1}{2} \tilde{C}_{12}^{(1)}
+ \frac{\beta-1}{2} \tilde{C}_{12}^{(2)}
-\beta (\tilde{C}_{22}^{(1)}+\tilde{C}_{22}^{(2)}) 
=0,
\nonumber\\
&
\beta (\tilde{C}_{11}^{(1)}+\tilde{C}_{11}^{(2)}) 
- \frac{\beta-1}{2} \tilde{C}_{21}^{(1)} 
- \frac{\beta+1}{2} \tilde{C}_{21}^{(2)}
=0,
\nonumber\\
&
\beta  (\tilde{C}_{12}^{(1)}+\tilde{C}_{12}^{(2)}) 
- \frac{\beta-1}{2} \tilde{C}_{22}^{(1)} 
- \frac{\beta +1}{2} \tilde{C}_{22}^{(2)}
=\tilde{r}.
\label{eq_Diophantine_TBG3}
\end{align} 
where $\beta =(1/\sqrt{3}) \cot (\theta/2)$.

In the TBGs with low-angle angles,
it is known that the nearly-flat bands around the charge neutral point
are separated from the rest of the spectrum by energy gaps. \cite{cao2016superlattice,cao2018unconventional,cao2018mott,nam2017lattice,koshino2018maximally}
If we assume that the Fermi energy lies in the gap just above the flat band, for example,
we have $\tilde{r} =4$ (relative to the charge neutral) by including the spin and valley degeneracies.
The set of equations of Eq. (\ref{eq_Diophantine_TBG3}) can be regarded as identities for a variable $\beta$,
as it stands for any $\theta$'s in the low-angle regime.
Then the Chern numbers are uniquely determined as
$\tilde{C}_{11}^{(1)}=\tilde{C}_{22}^{(1)}=4$, $\tilde{C}_{11}^{(2)}=\tilde{C}_{22}^{(2)}= -4$
and otherwise 0.
Actually, we can show that 
sliding the layer $l$ by $\mathbf{a}_{i}^{(l)}$
leads to the moir\'{e}-pattern movement by $\pm \mathbf{L}^M_{i}$ with $\pm$ for $l=1,2$, respectively.
The above solution of $\tilde{C}_{ij}^{(l)}$ means 
that four electrons trapped at each AA-stacking region precisely follow the movement of the moir\'{e} pattern,
as naturally expected.

To conclude,  we studied the topological charge pumping driven by interlayer sliding in the moir\'{e} superlattices.
The number of pumped charges is quantized to the sliding Chern numbers, which can be found
as a solution of a Diophantine equation. 
When the Fermi energy in the energy gap above or below the nearly-flat bands
of the twisted bilayer graphene, four electrons per a superlattice period 
are conveyed following the flow of the moir\'{e} pattern
perpendicularly to the sliding direction.
The interlayer sliding in moir\'{e} superlattices is, in principle, experimentally feasible by using a 
mechanical device \cite{ribeiro2018twistable}.
We expect that the slide-driven topological pumping may be observed as an electric current by
source and drain electrodes appropriately attached. If the system is isolated, 
the interlayer sliding should generate an electric polarization by accumulated charge at the edge.
This naturally implies the existence of the edge states in the energy gap with non-zero sliding Chern numbers.
The bulk-edge correspondence of this problem will be studied elsewhere. 
Also, the topological pumping in other moir\'{e} bilayer systems,
such as graphene/hBN  \cite{kindermann2012zero, wallbank2013generic, mucha2013heterostructures, jung2014ab, moon2014electronic,dean2013hofstadter,ponomarenko2013cloning,hunt2013massive,yu2014hierarchy}  
and transition metal dichalcogenides bilayers
\cite{wu2018hubbard,naik2018ultraflatbands,seyler2019signatures,tran2019evidence,jin2019observation,alexeev2019resonantly}, is left for future work.

The authors thank fruitful discussions with P.\ Kim and J.\ C.\ Hone. 
The authors are supported by JSPS KAKENHI Grant Number JP17K05496.

\bibliography{moire_pump}

\end{document}